\documentclass[sigconf]{acmart}

\usepackage{amsmath}

\AtBeginDocument{
  \providecommand\BibTeX{{
    \normalfont B\kern-0.5em{\scshape i\kern-0.25em b}\kern-0.8em\TeX}}
}

\acmConference[RLEM20]{Workshop on Reinforcement Learning for Energy Management in Buildings \& Cities}{November, 2020}{Virtual Meeting}

\copyrightyear{2020} 
\acmYear{2020} 
\setcopyright{usgovmixed}\acmConference[RLEM'20]{The 1st International Workshop on Reinforcement Learning for Energy Management in Buildings \& Cities}{November 17, 2020}{Virtual Event, Japan}
\acmBooktitle{The 1st International Workshop on Reinforcement Learning for Energy Management in Buildings \& Cities (RLEM'20), November 17, 2020, Virtual Event, Japan}
\acmPrice{15.00}
\acmDOI{10.1145/3427773.3427872}
\acmISBN{978-1-4503-8193-2/20/11}

\begin{document}

\title{A Comparison of Model-Free and Model Predictive Control for Price Responsive Water Heaters}

\author{David J. Biagioni, Xiangyu Zhang, Peter Graf, Devon Sigler, Wesley Jones}

\renewcommand{\shortauthors}{Biagioni, et. al.}

\begin{abstract}
We present a careful comparison of two model-free control algorithms, Evolution Strategies (ES) and Proximal Policy Optimization (PPO), with receding horizon model predictive control (MPC) for operating simulated, price responsive water heaters.  Four MPC variants are considered: a one-shot controller with perfect forecasting yielding optimal control; a limited-horizon controller with perfect forecasting; a mean forecasting-based controller; and a two-stage stochastic programming controller using historical scenarios.  In all cases, the MPC model for water temperature and electricity price are exact; only water demand is uncertain.  For comparison, both ES and PPO learn neural network-based policies by directly interacting with the simulated environment under the same scenarios used by MPC.  All methods are then evaluated on a separate one-week continuation of the demand time series.  We demonstrate that optimal control for this problem is challenging, requiring more than 8-hour lookahead for MPC with perfect forecasting to attain the minimum cost.  Despite this challenge, both ES and PPO learn good general purpose policies that outperform mean forecast and two-stage stochastic MPC controllers in terms of average cost and are more than two orders of magnitude faster at computing actions.  We show that ES in particular can leverage parallelism to learn a policy in under 90 seconds using 1150 CPU cores.
\end{abstract}


\maketitle
\section{Introduction}
In the modern smart grid, time-varying electricity prices, such as time-of-use (TOU) and real-time price (RTP), have been adopted in many electricity markets to incentivize end-user participation in demand response (DR) programs. Electric water heaters (EWH) are a good candidate for DR because of their prevalence and capacity to operate flexibly without jeopardizing user comfort. As a result, there have been a number efforts to facilitate EWH control: e.g., on EWH modeling for DR\cite{xu2014modeling}, optimal control of EWH under dynamic pricing \cite{lin2017optimal} and field verification on load reduction potential of EWH control \cite{heffner2005innovative}. Most existing EWH control algorithms are model-based and must directly confront nonlinear dynamics, stochasticity, system identification, and potentially high cost of numerical optimization. In contrast, some studies use reinforcement learning (RL) as an alternative, model-free control framework that naturally bypasses many of these challenges. Ruelens \textit{et al.} \cite{ruelens2016reinforcement} investigate using fitted Q-iteration to train an EWH control policy to minimize the operational cost under day-ahead price, and demonstrate a 15\% energy savings compared with the default controller. A similar study, also using fitted Q-iteration, considers EWH control in the context of local solar generation \cite{de2017using}. 

Still, important knowledge gaps remain.  First, in the existing literature, RL controllers typically compared with rule-based hysteresis or simple, deterministic MPC controllers, leaving the question open as to which approach is better suited for the task.  A thorough performance comparison of RL and MPC in the presence of uncertainty will help practitioners better evaluate trade-offs between model-free and model-based control.  Second, state-of-the art algorithms should be studied to provide an updated assessment for model-free control in terms of both control performance and computational cost. To help address these gaps, we conduct a careful comparison between two model-free controllers based on Evolution Strategies \cite{salimans2017evolution} and Proximal Policy Optimization \cite{schulman2017proximal}, and MPC controllers using both single-scenario forecasts and two-stage stochastic programs for the purpose of price responsive EWH control.

\section{Methodology}
\subsection{Modeling and Simulation}
\subsubsection{Electric Water Heater}
We adopt a single-element EWH model based on energy flow analysis as described in  \cite{dolan1992development,nehrir2007power} and use it for both simulation and as the dynamical model for MPC.  Water temperature $T$ in ${}^{\circ}$F is given by the discretized equation, 
\begin{align}
    T_{t+1} = T_t e^{-\Delta t/R'C} + \left( R'GT_{out} + R'T_{in}B_t + R'Q_t \right) \left(1 - e^{-\Delta t/R'C} \right),
    \label{temp-dynamics}
\end{align}
\noindent where $\Delta t$ is the length of the discrete time step and $R'GT_{out}$, $R'T_{in}B_t$, $R'Q_t$ are factors related to ambient temperature, inflow / outflow, and heating, respectively.  The quantities $R',G,C$ in (\ref{temp-dynamics}) are physical constants while $T_{out}, T_{in}$ are environmentally determined but here assumed to be constant. The quantity $B_t$ a is rational function in water demand, which we denote by $D_t$, while $Q_t$ is proportional to the heating element power output, denoted by $P_t$.  Power output takes on the discrete values of $\left\{ 0, P^{\text{on}} \right\}$ kW when off and on, respectively.  We refer the reader to \cite{nehrir2007power} for additional details.

In this study, water demand data is simulated using the STochastic Residential water End-use Model (STREaM)  \cite{stream,cominola2018implications}, using the default settings for a two-person household with faucet, shower, bathtub, clothes washer and dish washer.  Appliances such as toilets that do not consume hot water are excluded.

\subsection{Problem Formulation}
For the purpose of comparing model-free control with MPC, we formulate the EWH control problem as both a Markov Decision Process (MDP) and an equivalent optimization model cast as a mixed-integer linear program (MILP).  We begin by describing the reward function which is identical in both.

\subsubsection{Reward Function}
Our goal is to create an EWH controller that can co-optimize power cost (on- and off-peak) and discomfort associated with end-use consumption below a temperature threshold.  We denote this threshold, sometimes referred to as the ``cold shower" threshold, by $\underline{T}$.  The step reward we consider is given by
\begin{align}
    r_t = -\left( r_{p,t} P_t + r_c D_t T_{c,t}\right),
    \quad \text{where} \quad r_{p,t} =
    \begin{cases}
      \underline{r}_p, & \text{on-peak,}\\
      \overline{r}_p, & \text{off-peak}.\\
    \end{cases}
    \label{reward-fun}
\end{align}
\noindent Here $r_{c}$ is a cold shower discomfort cost and  $T_{c,t}=\operatorname{max}(0, \underline{T} - T_t)$ is the cold shower temperature, which only incurs a cost if the end-user is actively drawing water.  The second term in (\ref{reward-fun})  can be interpreted as a soft lower bound on user discomfort.

While we do not explicitly include a term in (\ref{reward-fun}) that models mechanical wear, this is an important issue automatically addressed by simple deadband control.  All controllers described here limit mechanical wear by instead enforcing a minimum down time for the EWH:  once the heating element has been turned off, it must stay off for a predefined amount of time.  The off duration can be chosen reasonably by analyzing the deadband control behavior itself; for the purposes of this paper, we use 10-minute minimum down time.

\subsubsection{Markov Decision Process}
We first formulate the EWH control problem as a Markov Decision Process (MDP):  $s\in\mathcal{S}$ denotes the state vector, $a\in\mathcal{A}$ the action vector, $r\in\mathcal{R}: \mathcal{S}\times\mathcal{A}\rightarrow\mathbb{R}$ the reward function, and $\pi\in\Pi: \mathcal{S}\rightarrow\mathcal{A}$ the stationary policy.  A probabilistic state transition function, $\rho: \mathcal{S}\times\mathcal{A} \rightarrow \mathcal{S}$, governs the underlying dynamics.

 A general MDP formulation of the EWH control problem is to find policy $\pi$ satisfying 
\begin{align}
    \underset{\pi\in\Pi}{\operatorname{max}} \quad & \mathbb{E}_{s_{t+1} \sim \rho(\cdot | s_t, a_t)} \left[ \sum_{t=0}^{\infty} \gamma^t r_t \left( s_t, a_t \right) \right]
        \label{mdp-objective}
\end{align}
with discount factor $0\leq\gamma\leq 1$.  In order to enable a direct comparison with MPC, we set $\gamma=1$ for the remainder of the discussion, and consider the finite-horizon, episodic case where the summation in (\ref{mdp-objective}) is taken over $\tau$ steps, $t=0,\dots,\tau-1$. The state vector is then defined to be $s_t=(T_t,T_{c,t},\chi_{t},D_t,\tilde{t})$ where $\chi_{t}$ is binary indicator of peak pricing (0: off-peak, 1: on-peak), and $\tilde{t}\in[0,1]$ is normalized time of day.  The  action is simple on/off control, $a_t = \pi(s_t)=P_t\in\left\{ 0, P^{\text{on}} \right\}$.

\subsubsection{Model Predictive Control}
Standard MPC formulations are similar to (\ref{mdp-objective}) with three typical exceptions: 1) the objective function is cast as a cost minimization over a lookahead horizon of length $K$, both to mollify the optimization problem and to discount the effects of increasing uncertainty; 2) decision variables correspond to a control sequence, $\{a_{k}\}_{k=0}^{K-1}$, rather than a policy $\pi$; 3) state transitions are modeled via a discrete dynamical system. Forecast-based MPC controllers handle uncertainty via feedback, i.e., by repeatedly re-solving the deterministic model with updated initial conditions and forecasts at each control step.  In this paper, we go one step further and formulate a MPC controller that solves a two-stage stochastic program at each control step to minimize expected cost (the negative of the objective in (\ref{mdp-objective})).  Because explicit evaluation of the expectation is not possible for most problems, it is commonly replaced by a sample mean over scenarios $n=1,...,N$, each with probability mass, $p^{(n)}$.  In this context, state variables evolve differently along each scenario which we denote by, e.g., $s^{(n)}_t$.  We note that this approach encompasses deterministic formulations that use forecasting, in particular, by using the forecast as the lone scenario.

The two-stage stochastic program is presented below as a mixed-integer linear program (MILP) which, in this paper, is exact with respect to the simulated dynamics.  In what follows, we use the shorthand $(\cdot)_k$ to mean $(\cdot)_{t_k}$ where $t_k = t + k\Delta t$ for time steps of size $\Delta t$, and explicitly set $a = P$ as the control action.  Simple bound constraints and initial conditions are omitted for space.
\begin{align}
    \underset{P_k}{\operatorname{min}} & \sum_{n=1}^N p^{(n)} \sum_{k=0}^{K-1} -\left( r_{p,k} P_{k} +  r_c T^{(n)}_{c,k} D^{(n)}_{k} \right),
        \label{milp-objective}\\
    \text{s.t.}& \nonumber\\
    T_k^{(n)} &= T_{k-1}^{(n)} e^{-c_1 \Delta t} + \left(c_2 + c_3(D_{k-1}^{(n)}) + c_4 P_{k-1} \right) \left(1 - e^{-c_1 \Delta t} \right),
        \label{milp-temp}\\
     \underline{T} &\leq (T_k^{(n)} + T_{c,k}^{(n)}),
        \label{milp-cs}\\
    T_k^{(n)}  &\leq \overline{T},
        \label{milp-temp-ub}\\
    y_{k-1} - y_{k}  &\leq z_{k},
    	\label{shutdown-1}\\
    y_{k} - y_{k-1} &\leq 1- \displaystyle\sum_{i = k-DT}^{k-1} z_{i}, \text{ for $k \geq DT$},
    	\label{shutdown-2}\\
    y_{k} - y_{k-1}  &\leq 1- \left(\displaystyle\sum_{i = 0}^{k-1} z_{i} + \displaystyle\sum_{i = k - DT}^{-1} \bar{z}_{i}\right), \text{ for $k < DT$},
    	\label{shutdown-3}\\
    P_k &\leq M y_k.
        \label{milp-force-off}
\end{align}
\noindent Here, $c_{(\cdot)}$ are physical constants collected in (\ref{temp-dynamics}) and $c_3(D_k^{(n)})$ indicates the parameter's dependence on water demand; $M$ is a large, positive constant ("big-M"); $DT$ is the number of time steps representing the minimum down time; $y_0,\dots,y_{K-1}$ and $\bar{z}_{-DT},\dots, \bar{z}_{-1}$ are binary variables. Constraint (\ref{milp-temp}) expresses the temperature dynamics, while (\ref{milp-cs}) defines $T_{c,k}^{(n)}$ via non-negative slack, and (\ref{milp-temp-ub}) represents a safety upper bound on the water temperature. Constraint (\ref{shutdown-2}) enforces bounds and constraints on non-negative power consumption variables $P_k$. The binary decision variables $y_{k}$ indicate on/off status of the heating element, and the binary variables  $z_{i}$, in conjunction with constraint (\ref{shutdown-1}), track when power consumption is switched on. Finally, constraints (\ref{shutdown-2}) - (\ref{milp-force-off}) that if power is switched off it cannot be switched back on for $DT$ time steps. 

\begin{figure}[t!]
  \centering
  \includegraphics[width=\columnwidth]{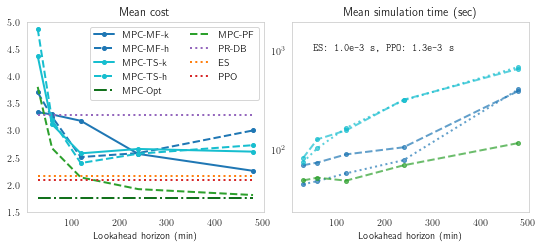}
  \caption{Comparison of mean controller performance.}
  \Description{}
  \label{fig:compare-mpc-cost}
\end{figure}

\subsection{Algorithms}
\subsubsection{Model Predictive Control}
All MPC optimization models described in this paper are MILP's and thus can be solved using robust open source and commercial solvers.  Reported results were obtained using Gurobi \cite{gurobi} with default settings.
\subsubsection{Model-Free Algorithms}
We consider two model-free algorithms, Evolution strategies (ES) \cite{salimans2017evolution,es2017blog} and Proximal Policy Optimization (PPO) \cite{schulman2017proximal,ppo2017blog}, applied to the EWH control problem.  The fundamentally differing approaches taken by these algorithms to model-free control makes them attractive subjects of study in this context. ES is a black-box optimization algorithm that performs a direct search on the policy space to maximize expected cumulative reward, rather than using backpropagation as is typically needed in policy gradient- or value-based RL.  This leads to a number of advantages including straightforward parallelization to 1000's of CPU cores and robustness to long episode horizons.  PPO, in contrast, is a popular actor-critic method that is relatively easy to implement and has proven empirically effective in many RL problems.  While PPO can also be scaled up to many cores, the use of backpropagation introduces a need for all-to-all communication of  gradient updates leading to an intrinsic computational bottleneck that is not present with ES.
\section{Results and Discussion}
\subsection{Numerical Studies}
We conducted numerical experiments comparing the performance of ES, PPO, MPC with perfect forecasting (MPC-PF), MPC with mean-forecasting (MPC-MF), and two-stage stochastic MPC (MPC-TS).  Price responsive deadband (PR-DB) control, which uses simple deadband control off-peak and remains off during peak, served as a simple baseline.  To provide an optimal baseline, we solved MPC-PF over the entire horizon as a one-shot optimization problem (MPC-Opt): because the physics is perfectly represented in the MPC models, and there is no demand uncertainty under perfect forecasting, MPC-Opt represents a truly optimal baseline.

We consider the problem of using $N=7$ scenarios for all methods, either as historical data (the previous week) or as representative for the previous three weeks, by performing k-means clustering and taking as representatives the scenarios closest to each cluster centroid.  For MPC-MF, scenarios are averaged to generate a single forecast scenario; for MPC-TS, the scenarios are used outright in the stochastic program.  Both ES and PPO utilized the scenarios for training their respective policies.  All methods were evaluated against one week of data corresponding to the (previously unseen) continuation of the training set.

STREaM water demand traces were aggregated to 1-minute intervals for the month of March.  Ignoring appliances that do not use hot water, the daily simulated consumption for the household range between 7.9 and 208.2 gallons, with mean, median and standard deviation of 69.3, 66.8, and 47.2, respectively. 
\begin{figure}[t!]
  \centering
  \includegraphics[width=\columnwidth]{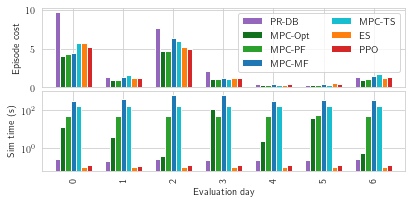}
  \caption{Controller performance on evaluation days.}
  \Description{}
  \label{fig:compare-control}
\end{figure}

All experiments used values of $\underline{r}_p=0.1$, $\overline{r}_p=1.0$, and $r_{c}=0.5$.  The ratio of $\overline{r}_p/\underline{r}_p=10$ is roughly in line with industry values for TOU prices \cite{sdge2020cpp}.  We note that the choice of $r_{c}$ coincides with a discomfort cost that equals on-peak cost when water temperature falls $9 ^\circ$F below the threshold $\underline{T}=115 ^\circ$F.  On-peak prices occur each day from 2-7pm, and all power consumed during that time incurs a per-unit cost of $\overline{r}_p$.  All controllers have perfect information about price.

Control sequences for MPC were generated using lookahead horizons of $0.5, 1, 2, 4$ and $8$ hours at 1-minute resolution.  In MPC-TS we set all scenarios to be equiprobable with $p^{(n)}=1/N$.  All methods used 10-minute control intervals ($DT=10$ in (\ref{milp-objective})-(\ref{shutdown-3})) during which the on/off control input cannot be changed, to prevent rapid switching.  This choice of minimum down time is empirically similar to the value under deadband control under 25 gal/hr constant demand; it also reduces the complexity of the MPC optimization problem considerably by decoupling the decision windows.

ES and PPO policies were computed using the RLLib library \cite{liang2017rllib}. Training for both algorithms was parallelized across 71, 143, 286, 574, and 1150 CPU cores; the best-performing policies discussed in the next section  were computed in 89 sec. (ES) and 64 min. (PPO) on 1150 and 287 CPU cores, respectively.
\begin{figure}[t!]
  \centering
  \includegraphics[width=\columnwidth]{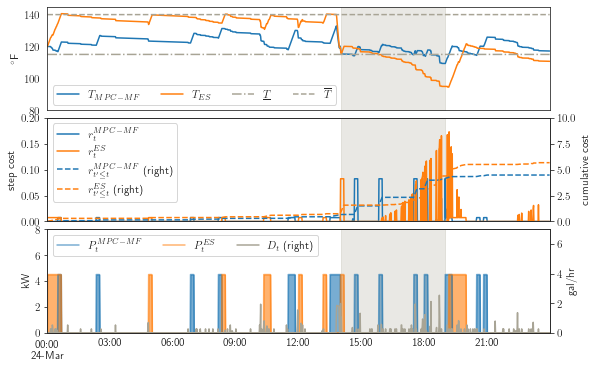}
  \caption{Comparison of ES with MP-MF for a single evaluation day in which MPC-MF outperforms ES.}
  \Description{}
  \label{fig:eval-24}
\end{figure}

\subsection{Results}
Figure \ref{fig:compare-mpc-cost} shows the mean performance of MPC methods on the 7 evaluation days, as a function of lookahead time.  Suffixes ``-k" and ``-h" indicate that scenarios are $k$-means centers and historical (most recent), respectively.  MPC-PF is seen to approach MPC-Opt with an 8-hour lookahead, illustrating the difficulty of achieving optimal control: the only uncertainty for MPC-PF is uncertainty beyond the lookahead horizon. MPC-MF performs best with 8-hour lookahead using k-means, while MPC-TS performs best with 2-hour lookahead using historical scenarios.  The right panel of Figure \ref{fig:compare-mpc-cost} shows the mean CPU time to run one forward simulation, the cost of which is dominated by the need to compute actions.  We can infer exponential scaling in the MILP solution times with respect to the horizon length which is to be expected for off-the-shelf MILP solvers.  ES and PPO, on the other hand, are at least two orders of magnitudes faster than MPC methods.

Figure \ref{fig:compare-control} gives a more detailed view of controller performance on all evaluation days in terms of both episode cost and simulation time. Performance for non-baseline methods is reported only for the model with the lowest objective cost for each category.

\begin{figure}[t!]
  \centering
  \includegraphics[width=\columnwidth]{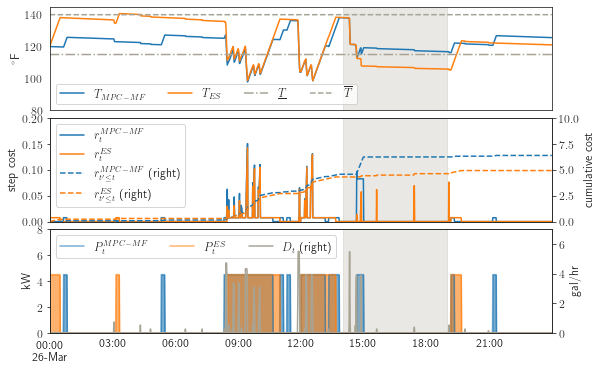}
  \caption{Comparison of ES with MP-MF for a single evaluation day in which ES outperforms MPC-MF.}
  \Description{}
  \label{fig:eval-26}
\end{figure}

Figures \ref{fig:eval-24} and \ref{fig:eval-26} provide a more detailed comparison of the control trajectories for ES and MPC-MF by plotting system dynamics, costs, and control inputs for two high-cost evaluation days (days 0 and 2 in Figure \ref{fig:compare-control}, respectively).  ES appears to learn a general strategy of pre-heating before peak pricing and to stay off during the peak; and reacts to demand more or less aggressively depending on the current price of power.  This is an effective strategy when, as in Figure \ref{fig:eval-26}, demand spikes early in the day but is relatively low on-peak.  MPC-MF, in this case, suffers from poor prediction; it turns the heater around 15:00 in anticipation of cold showers that never appear.  In Figure \ref{fig:eval-24}, the demand pattern consists of longer periods of low demand that reach their highest values towards the end the price peak.  We hypothesize that this is a historically likely occurrence because MPC-MF heats the water just in time to get past the worst of the cold shower costs (around 18:00).  ES, on the other hand, reacts to, rather than predicts, this occurrence and thus performs slightly worse during peak pricing.

\section{Conclusion}
We have demonstrated that model-free RL methods are at least viable candidates versus state-of-the-art MPC for EWH price response under demand uncertainty.  This is a case where occasional failure is not a huge disaster:  unlike unsupervised operation of the power grid itself, for example, a few cold showers, or a few high priced outlier days, may be acceptable if the mean cost over time is minimized.  Beyond the mere demonstration of the effectiveness of model-free control, we highlight several points for future study.

First, there is an inherent asymmetry regarding the uncertainty of future demand, namely, that its underestimation can lead to high costs on-peak, while overestimating has no immediate advantage.  This fact illustrates the need for storage to make such systems more effective by, e.g., allowing unused power to be sold back to the grid.  We see this as an important step to maximizing the impact of smart, price responsive EWH.

Second, for Figure \ref{fig:eval-24}, we noted that the fact that MPC-MF (recall ``MF" is the mean forecast case) works well in this case implies that this is a fairly ``typical" day, i.e., one well described by an average forecast.  ES, PPO and MPC-TS, in contrast, aim to maximize the mean reward over all scenarios.  This observation highlights the important fact that ``\emph{mean reward over scenarios} does not equal \emph{reward over mean scenario}".  And even here, not all scenarios are created equal:  those used in this paper led to good outcomes for model-free methods but not for MPC-TS.  We hope to explore the importance of scenario generation for both model-free and model-based EWH control in future studies.

Third, model-free algorithms such as ES can directly leverage parallelism on a simulated environment that is fast to evaluate, and for which there are no consequences for catastrophic decisions.  And, indeed, growing emphasis on creating ``digital twins" for energy systems makes simulation-based learning a viable option.  But how to best translate such policies back to the physical twin is not always clear.  This question, as well as the role that model uncertainty itself plays in control performance (for RL and MPC), are important subjects for future work.

Finally, there is increasing interest in combining elements of model-free and model-based control by, e.g., creating MPC models around ``learned" dynamics, or allowing an RL agent to use existing dynamics models to improve planning and accelerate learning.  We believe this is a promising research direction and hope to extend our study to include such hybrid methods.

\begin{acks}
This work was authored by the National Renewable Energy Laboratory, operated by Alliance for Sustainable Energy, LLC, for the U.S. Department of Energy (DOE) under Contract No. DE-AC36-08GO28308. Funding provided by the Assessment of Reinforcement Learning for Model NREL Problems Project and the Autonomous Energy Systems project, both funded by the National Renewable Energy Laboratory’s Laboratory Directed Research and Development program. The views expressed in the article do not necessarily represent the views of the DOE or the U.S. Government. The U.S. Government retains and the publisher, by accepting the article for publication, acknowledges that the U.S. Government retains a nonexclusive, paid-up, irrevocable, worldwide license to publish or reproduce the published form of this work, or allow others to do so, for U.S. Government purposes.
\end{acks}

\bibliographystyle{ACM-Reference-Format}
\bibliography{refs}










\end{document}